\global\def\draftcontrol{0}

   \def\versionno{ rotation }
\catcode`\@=11

\expandafter\ifx\csname draftcontrol\endcsname\relax\global\def\draftcontrol{0}
\fi

{\count255=\time\divide\count255 by 60
\xdef\hourmin{\number\count255}
\multiply\count255 by-60\advance\count255 by\time
\xdef\hourmin{\hourmin:\ifnum\count255<10 0\fi\the\count255}}
\def\draftdate{\number\month/\number\day/\number\year\ \ \ \hourmin }

\newcommand\makepapertitle{\par
  \begingroup
    \renewcommand\thefootnote{\@fnsymbol\c@footnote}%
    \def\@makefnmark{\rlap{\@textsuperscript{\normalfont\@thefnmark}}}%
    \long\def\@makefntext##1{\parindent 1em\noindent
            \hb@xt@1.8em{%
                \hss\@textsuperscript{\normalfont\@thefnmark}}##1}%
     \newpage
     \global\@topnum\z@   % Prevents figures from going at top of page.
     \@makepapertitle
     \thispagestyle{empty}\@thanks
  \endgroup
  \setcounter{footnote}{0}%
  \global\let\thanks\relax
  \global\let\makepapertitle\relax
  \global\let\@makepapertitle\relax
  \global\let\@thanks\@empty
  \global\let\@author\@empty
  \global\let\@date\@empty
  \global\let\@title\@empty
  \global\let\title\relax
  \global\let\author\relax
  \global\let\date\relax
  \global\let\and\relax
  \def\version{\let\version\@version\@gobble}
}
\def\@makepapertitle{%
  \newpage
   \ifnum\draftcontrol=1 {}
   \version\versionno
   \vskip 3em%
   \else
   \hfill\hbox to 3cm {\parbox{4cm}{\@pubnum}\hss}%
   \vskip 3em%
   \fi
   \begin{center}%
   \let \footnote \thanks
     {\LARGE {\@title}}%
     \vskip 1.5em%
     {\normalsize%\large
       \lineskip .5em%
       \begin{tabular}[t]{c}%
         \@author
       \end{tabular}\par}%
     \vskip 1.5em%
     {\@bstract}%
     \end{center}%
     \vskip 1.5em
     \@date%
   \par
}

\gdef\@pubnum{}
\def\pubnum#1{%
  \gdef\@pubnum{#1}}

\gdef\@bstract{}
\def\Abstract#1{%
  \gdef\@bstract{%
   \parbox{\textwidth-0pc}{%
   \centerline{\bf Abstract}\penalty1000%
\kern.2cm%
\noindent%\abstractfont \baselineskip=12pt
\renewcommand\baselinestretch{1.0}%
{#1}}}
}

\def\ps@paper{\let\@mkboth\@gobbletwo%
     \ifnum\draftcontrol=1
    \def\@oddfoot{\hbox to \textwidth{\tiny \versionno \hfil\tiny\draftdate}%
    \hskip -\textwidth \hbox to \textwidth{\hfil\rm\thepage\hfil}}%
     \else\def\@oddfoot{\hbox to \textwidth{\hfil\rm\thepage\hfil}}
     \fi
     \let\@evenfoot\@oddfoot
}
\def\body{\clearpage
          \pagestyle{paper}
    }
\def\@version#1{\ifnum\draftcontrol=1
\typeout{}\typeout{#1}\typeout{}
\vskip3mm\centerline{\hbox{\fbox{\normalsize{\tt DRAFT -- #1 -- }
                   {\draftdate}}}}\vskip3mm
\fi}
\let\version\@version
\long\def\eqlabel#1{\ifnum\draftcontrol=1
                    \tag@false  % there are some problems with multline without this
                    \tag*{(\theequation) \hbox to -0.2cm{\hspace{0cm}\small{#1}\hss}}
                    \refstepcounter{equation}
                    \edef\@currentlabel{\theequation}
                    \ltx@label{#1}          % use old LaTeX \label instead of new definition
                    \else
                    \label{#1}
                    \fi
                    }
\let\st@bibitem\@bibitem
\let\st@lbibitem\@lbibitem
\ifnum\draftcontrol=1
  \def\@bibitem#1{%
    \st@bibitem{#1}\a@@label{#1}\ignorespaces}
  \def\@lbibitem[#1]#2{%
    \st@lbibitem[#1]{#2}\a@@label{#2}\ignorespaces}
  \def\a@@label#1{%
    \gdef\a@lab{\smash{\normalfont\small#1}}
    \ifvmode
      \if@inlabel
        \global\setbox\@labels\hbox{%
          \llap{\a@lab\let\a@lab\relax
                \kern\@totalleftmargin\kern\marginparsep}%
          \box\@labels}%
      \fi
    \fi}
\fi
\documentclass[12pt,letterpaper]{article}
\usepackage{amsmath,amssymb,array,calc,rotating,epsfig,psfrag}
\usepackage[nosort]{cite}
\ifnum\draftcontrol=1
\fi
\renewcommand\baselinestretch{1.25}
\renewcommand\section{\@startsection {section}{1}{\z@}%
                                   {-3.5ex \@plus -1ex \@minus -.2ex}%
                                   {2.3ex \@plus.2ex}%
                                   {\normalfont\large\bfseries}}
\renewcommand\subsection{\@startsection{subsection}{2}{\z@}%
                                   {-3.25ex\@plus -1ex \@minus -.2ex}%
                                   {1.5ex \@plus .2ex}%
                                   {\normalfont\normalsize\bfseries}}
\renewcommand\subsubsection{\@startsection{subsubsection}{3}{\z@}%
                                   {-3.25ex\@plus -1ex \@minus -.2ex}%
                                   {1.5ex \@plus .2ex}%
                                   {\normalfont\normalsize\it}}
\renewcommand\paragraph{\@startsection{paragraph}{4}{\z@}%
                                   {-3.25ex\@plus -1ex \@minus -.2ex}%
                                   {1.5ex \@plus .2ex}%
                                   {\normalfont\normalsize\bf}}

\numberwithin{equation}{section}

\def\ie{{\it i.e.}}

\def\revise#1       {\raisebox{-0em}{\rule{3pt}{1em}}%
                     \marginpar{\raisebox{.5em}{\vrule width3pt\
                     \vrule width0pt height 0pt depth0.5em
                     \hbox to 0cm{\hspace{0cm}{%
                     \parbox[t]{4em}{\raggedright\footnotesize{#1}}}\hss}}}}

\def\calc         {{\cal C}}

\def\calf         {{\cal F}}

\def\call         {{\cal L}}
\def\calm         {{\cal M}}
\def\caln         {{\cal N}}
\def\calo         {{\cal O}}

\def\complex      {{\mathbb C}}

\def\zet          {{\mathbb Z}}

\def\del          {\partial}

\def\tr           {\mathop{\rm Tr}}

\def\sqr#1#2{{\vcenter{\vbox{\hrule height.#2pt
 \hbox{\vrule width.#2pt height#1pt \kern#1pt
 \vrule width.#2pt}\hrule height.#2pt}}}}

\newcommand{\fft}[2]{{\frac{#1}{#2}}}
\newcommand{\ft}[2]{{\textstyle{\frac{#1}{#2}}}}

\def\a{\alpha}
\def\b{\beta}

\def\w{\omega}

\def\rp{r_+}

\newcommand{\ww}{\mathfrak{w}}

\def\w{\omega}

\def\a{\alpha}

\def\tq{\tilde{q}}
\def\p{\phi}

\catcode`\@=12

\begin{document}

%%%
%%%%%% text starts here
%%%%%%%%%

\title{Gauged supergravity from type IIB string theory\\[6pt]
on $Y^{p,q}$ manifolds}

\pubnum{%
UWO-TH-06/11\\
MCTP-06-18}
\date{July 2006}

\author{
Alex Buchel$^{1,2}$ and James T. Liu$^{3}$\\[0.4cm]
\it $^1$Perimeter Institute for Theoretical Physics\\
\it Waterloo, Ontario N2J 2W9, Canada\\[0.1cm]
\it $^2$Department of Applied Mathematics\\
\it University of Western Ontario\\
\it London, Ontario N6A 5B7, Canada\\[0.1cm]
\it $^3$Michigan Center for Theoretical Physics\\
\it Randall Laboratory of Physics, The University of Michigan\\
\it Ann Arbor, MI 48109--1040, USA
}

\Abstract{
We first construct a consistent Kaluza-Klein reduction ansatz for
type IIB theory compactified on Sasaki-Einstein manifolds $Y^{p,q}$ with
Freund-Rubin 5-form flux giving rise to minimal $\caln=2$ gauged
supergravity in five dimensions.  We then investigate the $R$-charged
black hole solution in this gauged supergravity, and in particular
study its thermodynamics.  Based on the gauge theory/string theory
correspondence, this non-extremal geometry is dual to finite temperature
strongly coupled four-dimensional conformal gauge theory plasma with
a $U(1)_R$-symmetry charge chemical potential.  We study transport
properties of the gauge theory plasma and show that the ratio of
shear viscosity to entropy density in this plasma is universal. We
further conjecture that the universality of shear viscosity of
strongly coupled gauge theory plasma extends to nonzero $R$-charge
chemical potential.
}

%\enlargethispage{1.5cm}

\makepapertitle

\body

\version\versionno

%%%%%%%%%%%%%%%%%%%%%%%%%%%%%%%%%%%%%%%%
\section{Introduction}

The gauge theory/string theory correspondence of Maldacena
\cite{m9711,Gubser:1998bc,Witten:1998qj,m2} provides a valuable insight
into the nonperturbative
dynamics of strongly coupled gauge theory plasma.  However, on the
string theory side, this duality is tractable only for weakly curved
backgrounds and with small string coupling, \ie, in the supergravity
approximation.  Following the Maldacena correspondence, this implies
that the dual four dimensional gauge theory must be strongly coupled
at all energy scales.  Unfortunately, this latter condition excludes
a dual supergravity description of real QCD.  Nonetheless, one
expects that certain universal properties of strongly coupled gauge
theory plasma derived in the planar limit and for large 't Hooft
coupling would persist for real QCD as well.

A noteworthy example of such a universal property is the ratio of shear
viscosity to the entropy density. In \cite{u1,u2,u3} a theorem was proven
that, in the absence of chemical potentials, a gauge theory plasma in the
planar limit and for infinitely large 't Hooft coupling%
\footnote{Finite 't Hooft coupling corrections to shear viscosity of the
$\caln=4$ supersymmetric $SU(N)$ Yang-Mills theory plasma were discussed
in \cite{f1,f2}.}
in theories that admit a dual supergravity description has a universal ratio
of shear viscosity $\eta$ to the entropy density $s$:
\begin{equation}
\frac{\eta}{s}=\frac{1}{4\pi}.
\eqlabel{univ}
\end{equation}
Furthermore, other transport properties of the strongly coupled non-conformal
gauge theory plasma, like the speed of sound $c_s$ and the bulk viscosity
$\zeta$, while not universal, have a certain generic behavior in the
near-conformal limit \cite{nc1,nc2,nc3}
\begin{equation}
\frac{\zeta}{\eta}=-\kappa \left(c_s^2-\frac 13\right)
+\calo\left(\left(c_s^2-\frac 13\right)^2\right),\qquad \kappa\sim 2\dots5.
\eqlabel{xieta}
\end{equation}

Introduction of a nonzero chemical potential in the strongly coupled gauge
theory plasma violates the sufficient condition under which the universal
result \eqref{univ} has been derived \cite{u1,u3}.  On the other hand, it
was recently shown \cite{r1,r2,r3} that the universality \eqref{univ}
persists in $\caln=4$ supersymmetric Yang-Mills plasma even with nonzero
$R$-charge chemical potentials%
\footnote{The shear viscosity of the M2-brane plasma also appears to
satisfy the universal relation \eqref{univ} \cite{r4}.}.
Thus it appears that the shear viscosity relation \eqref{univ} is more robust
than originally anticipated in \cite{u1,u2,u3}.  However, further exploration
of shear viscosity with chemical potential in strongly coupled four-dimensional
gauge theory plasma has been hampered by the fact that, until recently, few models
exist where the dual supergravity solutions are easily accessible.

The most well understood system is of course $\caln=4$ Yang-Mills plasma,
which is holographically dual to IIB on AdS$_5\times S^5$.  From a
five-dimensional point of view, this dual is given by $\caln=8$
gauged supergravity, which may be obtained by reduction of the ten-dimensional
theory on $S^5$.  The $\caln=4$ Yang-Mills theory has an $SO(6)$ $R$-symmetry
which may correspondingly be identified with the isometries of $S^5$.  It
is then possible to turn on up to three commuting $R$-charges under the
maximum abelian subgroup $U(1)^3\subset SO(6)$.  At the same time, this
sub-sector of the full gauged supergravity admits a consistent truncation to
$\caln=2$ supergravity coupled to two abelian vectors, commonly referred to
as the STU model.  As a result, explicit investigations of the nonzero
$R$-charge sector of thermal $\caln=4$ Yang-Mills can be easily performed
in the context of the STU model.  In particular, non-extremal $R$-charged black
holes were constructed in \cite{stu1}, and their thermodynamics were
investigated in \cite{stu20,stu2,stu3,stu4}.  In addition, these
five-dimensional solutions can be uplifted to the full IIB theory and
investigated from the ten-dimensional point of view \cite{stu5,stu6}.

The uplifting of five-dimensional gauged supergravity solutions to ten
dimensions is closely related to the existence of a full non-linear
Kaluza-Klein reduction ansatz of the original IIB theory.  As it turns
out, a complete proof of the consistency of the reduction of IIB theory
on $S^5$ has yet to be given.  Nevertheless, it is generally assumed to
be consistent based on the linearized construction as well as explicit
reductions to subsets of the full $\caln=8$ theory in five dimensions.
In particular, the reduction to the STU model is given in \cite{stu5},
while reduction to $\caln=4$ in five dimensions was given in
\cite{kk1}, and reduction of the metric and self-dual 5-form on $S^5$
was given in \cite{kk2}.

Although $\caln=4$ Yang-Mills has proven to be a fruitful system to
investigate, it would be highly desirable to explore models with reduced
supersymmetry.  In this context, the explicitly constructed Sasaki-Einstein
metrics $Y^{p,q}$ \cite{y1,y2} have led to rapid new developments in
understanding their holographically dual $\caln=1$ Yang-Mills theories.
At the same time, however, much of this progress has been restricted to
either zero temperature or zero chemical potential (or both).  What we aim
to do below is to make progress towards the study of $\caln=1$
Yang-Mills plasma at finite $R$-charge chemical potential.  To do so, we
first construct the consistent non-linear Kaluza-Klein reduction of IIB on
$Y^{p,q}$ giving rise to gauged $\caln=2$ supergravity in five dimensions,
and then turn to the thermodynamics and hydrodynamics of the corresponding
$R$-charged black holes.

In general, the Kaluza-Klein reduction of a higher-dimensional theory on
an internal manifold $Y$ generally involves both `massless' lower-dimensional
fields (related to zero modes on $Y$) and the Kaluza-Klein tower of massive
states.  This distinction, however, is not entirely clear, except perhaps in
the case where $Y$ is a torus.  Consistency of the reduction to the massless
sector then depends on the decoupling of the Kaluza-Klein tower, and this
is a highly non-trivial condition.  For $T^n$ reductions, truncation to the
zero modes is guaranteed to be consistent, as this is a truncation to the
charge singlet sector of $U(1)^n$.  However, for spheres or more complicated
manifolds $Y$, it is often the case that the states that are considered
`massless' (such as states in the lower-dimensional gauged supergravity
multiplet) carry non-trivial internal charge.  These charged states then
have the potential of acting as sources to the Kaluza-Klein tower, thus
rendering the truncation inconsistent unless some appropriate symmetry or
conspiracy among fields is in place.

As an example of this inconsistency, consider for example the compactification
of IIB on $T^{1,1}$.  The resulting five-dimensional theory is $\caln=2$
gauged supergravity with a full Kaluza-Klein spectrum which was obtained
in \cite{kkt11}.  Because the isometry of $T^{1,1}$ is $SU(2)\times SU(2)
\times U(1)$, the massless gauge bosons transform under the identical group.
Since the $U(1)$ gauge boson is clearly the $\caln=2$ graviphoton coupling
to $U(1)_R$, the massless sector may be described as $\caln=2$ gauged
supergravity coupled to $SU(2)\times SU(2)$ vector multiplets.  It turns
out, however, that it is inconsistent to retain these $SU(2)\times SU(2)$
vector multiplets in any truncation to the massless sector \cite{hmp};
the only consistent truncation is to pure $\caln=2$ gauged supergravity.
This, however, is sufficient for our needs, as the pure supergravity
sector is all that is needed for uplifting the $R$-charged configurations
of interest.

The more general case of IIB reduced on $Y^{p,q}$ is similar.  For generic
$Y^{p,q}$, its isometry is $SU(2)\times U(1)\times U(1)$, and the resulting
Kaluza-Klein reduction yields $\caln=2$ gauged supergravity coupled to
$SU(2)\times U(1)$ vector multiples in the massless sector.  Only the
truncation to minimal $\caln=2$ gauged supergravity is consistent, and the
resulting geometries are holographically dual to certain strongly coupled
superconformal quiver gauge theories \cite{y3,y4}.  Working at finite
temperature and $R$-charge in these gauge theories corresponds to taking an
$R$-charged black hole in the supergravity description.  Such black
holes carrying $U(1)_R$ graviphoton charge have been discussed in \cite{stu1},
and simply correspond to STU black holes with three equal charges.
Given these black holes and the Kaluza-Klein reduction ansatz on $Y^{p,q}$,
we may then study the thermodynamics and hydrodynamics of such objects.
In particular, we examine the shear viscosity and the speed of sound in the
strongly coupled superconformal plasma which is holographically dual to
the black hole background.  We find that the shear viscosity and entropy
density in the quiver gauge theories \cite{y3,y4} continues to satisfy
the universality relation \eqref{univ}, even in the presence of a nonzero
chemical potential.  Although we do not provide a proof, we believe that
this universality \eqref{univ} is likely to be true in any gauge theory
plasma with $R$-charge chemical potentials turned on.

This paper is organized as follows.
In the following section, we present the Kaluza-Klein reduction ansatz
for IIB on $Y^{p,q}$ giving rise to minimal $\caln=2$ gauged supergravity
in five dimensions.  We then take up the thermodynamics of $R$-charged
black hole solutions of this $\caln=2$ gauged supergravity in section 3
and the hydrodynamics in section 4.  Finally, we present our conclusions
in section 5.

%%%%%%%%%%%%%%%%%%%%%%%%%%%%%%%%%%%%%%%%
\section{$\caln=2$ gauged sugra from type IIB string theory on $Y^{p,q}$
manifolds with 5-form fluxes}

A five-dimensional Sasaki-Einstein manifold $Y$ is an Einstein manifold
preserving some fraction of supersymmetry such that the cone over $Y$ is
(non-compact) Calabi-Yau.  It is well known that a Sasaki-Einstein manifold
$Y$ admits a constant norm Killing vector field, known as the Reeb vector.
The existence of the Reeb vector allows the metric on $Y$ to be written
as a $U(1)$ bundle over a four-dimensional K\"ahler-Einstein base $B$
\begin{equation}
ds^2(Y)=ds^2(B)+\ft19(d\psi+\mathcal A)^2,
\eqlabel{eq:reeb}
\end{equation}
where $d\mathcal A=6J$ and $J$ is the K\"ahler form on $B$.

The Freund-Rubin compactification of IIB on $S^5$ admits a straightforward
generalization where $S^5$ is replaced by the Sasaki-Einstein manifold $Y$.
Taking only the metric and self-dual 5-form, the only ten-dimensional IIB
equations of motion that we are concerned with are the Einstein and 5-form
equations
\begin{equation}
R_{MN}=\fft12\fft1{2\cdot4!}F_{MPQRS}F_N{}^{PQRS},\qquad
dF_{(5)}=0,\qquad
F_{(5)}=*F_{(5)}.
\end{equation}
In addition, the IIB gravitino variation is given by
\begin{equation}
\delta\psi_M=\left[\nabla_M+\fft{i}{16\cdot5!}F_{NPQRS}\Gamma^{NPQRS}\Gamma_M
\right]\epsilon.
\eqlabel{eq:10var}
\end{equation}
The resulting Freund-Rubin ansatz is then of the form
\begin{equation}
ds_{10}^2=ds^2(AdS_5)+ \fft1{g^2}ds^2(Y),\qquad
F_{(5)}=(1+*)G_{(5)},\qquad
G_{(5)}=4g\epsilon_{(5)},
\eqlabel{eq:fr}
\end{equation}
where $\epsilon_{(5)}$ is the five-dimensional volume form of AdS$_5$,
and $g$ is the coupling constant of the five-dimensional gauged supergravity
(corresponding to the inverse AdS$_5$ radius).  The five-dimensional
theory is described by $\caln=2$ gauged supergravity, and is holographically
dual to an $\caln=1$ superconformal gauge theory.

Since $Y$ has the form of (\ref{eq:reeb}), there is a natural symmetry
corresponding to the $U(1)$ fiber.  On the gravity side, the gauge boson
under this $U(1)$ is the $\caln=2$ graviphoton, while on the SCFT side this
symmetry is just the $U(1)_R$ symmetry.  In order to retain the graviphoton,
the Freund-Rubin metric ansatz (\ref{eq:fr}) may be extended in the obvious
manner
\begin{equation}
ds_{10}^2=g_{\mu\nu}dx^\mu dx^\nu+\fft1{g^2}\left(ds^2(B)+
\ft19(d\psi+\mathcal A+A_\mu dx^\mu)^2\right),
\eqlabel{eq:kkmet}
\end{equation}
where $g_{\mu\nu}$ is the five-dimensional metric and $A_\mu$ is the
graviphoton.  The reduction of $F_{(5)}$ has the form
\begin{equation}
F_{(5)}=(1+*)G_{(5)},\qquad
G_{(5)}=4g\epsilon_{(5)}-\fft1{3g^2}J_{(2)}\wedge *_5F_{(2)},
\eqlabel{eq:kkf5}
\end{equation}
where $J_{(2)}$ is the K\"ahler form on $B$ and $F_{(2)}=dA_{(1)}$ with
$A_{(1)}=A_\mu dx^\mu$.
At a linearized order 
in gauge potential $A_{\mu}$, and without the backreaction of its field strength on the metric,
the ansatz  \eqref{eq:kkmet}, \eqref{eq:kkf5} appeared previously in \cite{bhk}. We claim here that 
\eqref{eq:kkmet} and \eqref{eq:kkf5}  are in fact consistent at a  nonlinear level.

The above Kaluza-Klein reduction ansatz, (\ref{eq:kkmet}) and (\ref{eq:kkf5}),
gives rise to equations of motion that may be obtained from the
five-dimensional Lagrangian
\begin{equation}
e^{-1}\mathcal L_5=R+12g^2-\ft1{12}F_{\mu\nu}^2+\ft1{108}
\epsilon^{\mu\nu\rho\lambda\sigma}F_{\mu\nu}F_{\rho\lambda}A_\sigma.
\eqlabel{eq:5lag}
\end{equation}
This is the bosonic sector of minimal $\mathcal N=2$ gauged supergravity
in five dimensions.  As written, the graviphoton is non-canonically
normalized.  However canonical normalization may be achieved by the
rescaling $A_\mu\to\sqrt3A_\mu$.  Note, also, that the five-dimensional
Newton's constant $G_5$ of this $\caln=2$ gauged supergravity is related to
the ten dimensional Newton's constant $G_{10}$ according to
\begin{equation}
G_5=\frac{G_{10}}{{\rm volume}\left(Y\right) }.
\eqlabel{g5def}
\end{equation}

Although the above results hold in general, much of the interest in
this system is due to the availability of explicit Sasaki-Einstein metrics
$Y^{p,q}$ that were originally constructed in \cite{y1,y2}.  When IIB
is compactified on $Y^{p,q}$, the resulting AdS$_5$ backgrounds are
holographically dual to explicit $\caln=1$ superconformal quiver gauge theories
SCFT$^{p,q}$ \cite{y3,y4}.  The special case of $Y^{1,0}$ corresponds
to $T^{1,1}$, and the dual gauge theory was investigated in \cite{kw}.
To make the above more explicit, we will first work out the
$T^{1,1}$ reduction in detail, and then follow with the extension to
general $Y^{p,q}$ manifolds.

\subsection{$T^{1,1}$ reduction of IIB}

The well known case of IIB on AdS$_5\times T^{1,1}$ was originally
investigated in \cite{kw}, and this has led to numerous important variations.
Furthermore, this was one of only a few explicitly known examples of reduced
supersymmetry systems, at least until the recent construction of an infinite
family of $Y^{p,q}$ manifolds.  As indicated in the introduction, although
$T^{1,1}$ admits an $SU(2)\times SU(2)\times U(1)$ isometry, it is only
consistent to truncate to minimal $\caln=2$ supergravity with the $U(1)$
gauged by the $\caln=2$ graviphoton.  Following (\ref{eq:kkmet}) and
(\ref{eq:kkf5}), and expressing $T^{1,1}$ as $U(1)$ bundled over $S^2
\times S^2$, the reduction ansatz is as follows
\begin{eqnarray}
ds_{10}^2&=&g_{\mu\nu}dx^\mu dx^\nu+\fft1{g^2}\Bigl[
\ft16(d\theta_1^2+\sin^2\theta_1\,
d\phi_1^2)+\ft16(d\theta_2^2+\sin^2\theta_2\,d\phi_2^2)\nonumber\\
&&\qquad+\ft19(d\psi+\cos\theta_1\,d\phi_1+\cos\theta_2\,d\phi_2
+gA_{(1)})^2\Bigr],\nonumber\\
F_{(5)}&=&(1+*)\left[4g\epsilon_{(5)}+\fft1{18g^2}(\sin\theta_1\,d\theta_1
\wedge d\phi_1+\sin\theta_2\,d\theta_2\wedge d\phi_2)\wedge*_5F_{(2)}\right].
\label{eq:t11metf5}
\end{eqnarray}

Before proceeding with the reduction, it is natural to introduce a
vielbein basis for the $T^{1,1}$ coordinates
\begin{eqnarray}
&&e^1=\ft1{\sqrt6}d\theta_1,\qquad e^2=\ft1{\sqrt6}\sin\theta_1\,d\phi_1,\qquad
e^3=\ft1{\sqrt6}d\theta_2,\qquad e^4=\ft1{\sqrt6}\sin\theta_2\,d\phi_2,
\nonumber\\
&&e^5=\ft13(d\psi+\cos\theta_1\,d\phi_1+\cos\theta_2\,d\phi_2+gA_{(1)}).
\end{eqnarray}
Note that the self-dual K\"ahler form on the $S^2\times S^2$ base is
given by $J_{(2)}=-e^1\wedge e^2-e^2\wedge e^4$, where the sign is
chosen to match (\ref{eq:reeb}) and (\ref{eq:kkf5}).  Using the relations
\begin{eqnarray}
&&de^1=0,\qquad de^2=\sqrt6\cot\theta_1\,e^1\wedge e^2,\qquad
de^3=0,\qquad de^4=\sqrt6\cot\theta_2\,e^3\wedge e^4,\nonumber\\
&&de^5=2J_{(2)}+\ft13gF_{(2)},
\end{eqnarray}
we may obtain the spin-connections
\begin{eqnarray}
&&\omega^{12}=-\sqrt6\cot\theta\,e^2+e^5,\qquad
\omega^{34}=-\sqrt6\cot\theta_2\,e^4+e^5,\nonumber\\
&&\omega^{15}=e^2,\qquad\omega^{25}=-e^1,\qquad\omega^{35}=e^4,\qquad
\omega^{45}=-e^3,
\label{eq:sct11}
\end{eqnarray}
on $T^{1,1}$, as well as
\begin{equation}
\omega^{\alpha\beta}=\omega^{(5)\,\alpha\beta}-\fft1{6g}F^{\alpha\beta}e^5,
\qquad\omega^{\alpha5}=-\fft16F^\alpha{}_\beta e^\beta,
\eqlabel{eq:scmix}
\end{equation}
for the mixed components.

The above vielbeins are naturally given for the unit radius $T^{1,1}$.
Hence the rewriting of (\ref{eq:t11metf5}) in terms of the above vielbeins
introduces several factors of $g$ as follows:
\begin{eqnarray}
ds_{10}^2&=&g_{\mu\nu}dx^\mu dx^\nu+\fft1{g^2}\sum_i(e^i)^2,\nonumber\\
F_{(5)}&=&(1+*)\left[4g\epsilon_{(5)}-\fft1{3g^2}J_{(2)}
\wedge *_5F_{(2)}\right]\nonumber\\
&=&4\left(g\epsilon_{(5)}+\fft1{g^4}e^1\wedge e^2\wedge e^3\wedge e^4\wedge e^5
\right)-\fft1{3g^2}J_{(2)}
\wedge\left(*_5F_{(2)}+\fft1ge^5\wedge F_{(2)}\right).\nonumber\\
\label{eq:g5viel}
\end{eqnarray}
Since the K\"ahler form is closed, $dJ_{(2)}=0$, it is now easy to see that
the ten-dimensional Bianchi identity $dF_{(5)}=0$
yields the five-dimensional Bianchi identity and equation of motion for
$F_{(2)}$
\begin{equation}
dF_{(2)}=0,\qquad d*_5F_{(2)}=-\ft13F_{(2)}\wedge F_{(2)}.
\eqlabel{eq:5f}
\end{equation}
Note that, to verify the Bianchi identity, we also need the fact that
$J_{(2)}\wedge J_{(2)}$ gives twice the volume form on the Einstein-K\"ahler
base, namely $J_{(2)}\wedge J_{(2)}=2e^1\wedge e^2\wedge e^3\wedge e^4$.

Turning next to the ten-dimensional Einstein equation, we note that it
splits into several components, $\mu\nu$, $ij$, and $\mu i$.
We find that the $\mu\nu$ components gives the five-dimensional Einstein
equation
\begin{equation}
R_{\mu\nu}^{(5)}=-4g^2g_{\mu\nu}+\ft16(F_{\mu\lambda}F_\nu{}^\lambda
-\ft16g_{\mu\nu}F^2),
\eqlabel{eq:5eins}
\end{equation}
while the internal $ij$ components are identically satisfied.  In addition,
the mixed $\mu 5$ component yields the $2$-form equation of motion
\begin{equation}
\nabla^\lambda F_{\lambda\mu}=-\ft1{12}
\epsilon_{\mu\nu\rho\lambda\sigma}F^{\nu\rho}F^{\lambda\sigma},
\end{equation}
which is identical to the one given in (\ref{eq:5f}) in form notation.
As indicated at the beginning of this section, the combined equations of
motion (\ref{eq:5f}) and (\ref{eq:5eins}) may be obtained from the
Lagrangian (\ref{eq:5lag}) for minimal gauged supergravity in five dimensions.

Although we have only focused on the bosonic fields of the reduction, it
ought to be apparent that the full Kaluza-Klein reduction onto $\caln=2$
gauged supergravity including the gravitino is a consistent one.  To see
this, we may highlight the reduction of the IIB gravitino variation
(\ref{eq:10var}).  In order to proceed, we first introduce a decomposition
of the ten-dimensional Dirac matrices
\begin{equation}
\Gamma^A=\{\gamma^\alpha\otimes1\otimes\sigma^1,1\otimes\tilde\gamma^a\otimes
\sigma^2\}.
\end{equation}
By convention, we take the product of spacetime Dirac matrices to
be $\gamma^0\gamma^1\gamma^2\gamma^3\gamma^4=-i$ and internal Dirac
matrices to be $\tilde\gamma^1\tilde\gamma^2\tilde\gamma^3\tilde\gamma^4
\tilde\gamma^5=1$.  In addition,
the IIB gravitino $\psi_\mu$ and supersymmetry parameter $\epsilon$ are
ten-dimensional Weyl spinors satisfying the chirality projection
$\Gamma^{11}\epsilon=\epsilon$ where $\Gamma^{11}=\Gamma^0\ldots\Gamma^9
=1\otimes1\otimes\sigma^3$.  As a result, such chiral IIB spinors may be
written as
\begin{equation}
\epsilon=\varepsilon\otimes\eta\otimes\left[\begin{matrix}1\cr0\end{matrix}
\right].
\end{equation}

Given this decomposition of the Dirac matrices, we now proceed with the
reduction.  The gravitino variation (\ref{eq:10var}) is perhaps most
straightforwardly investigated in tangent space.  Using (\ref{eq:g5viel})
for the 5-form, and taking chirality into account, we note that
\begin{eqnarray}
F_{BCDEF}\Gamma^{BCDEF}\Gamma_\alpha&=&-40i[24g\gamma_\alpha-(\tilde\gamma^{12}
+\tilde\gamma^{34})F_{\beta\gamma}\gamma^{\beta\gamma}\gamma_\alpha,
\nonumber\\
F_{BCDEF}\Gamma^{BCDEF}\Gamma_a&=&40[24g\gamma_a-(\tilde\gamma^{12}
+\tilde\gamma^{34})F_{\beta\gamma}\gamma^{\beta\gamma}\tilde\gamma_a].
\label{eq:fdotgamma}
\end{eqnarray}
Combining this with the covariant derivatives formed from the spin connections
(\ref{eq:sct11}) and (\ref{eq:scmix}), we obtain the spacetime variation
\begin{equation}
\delta\psi_\alpha=[\nabla_\alpha^{(5)}-gA_\alpha\partial_\psi
+\ft{i}{12}F_{\alpha\beta}\gamma^\beta\tilde\gamma^5
-\ft1{48}F_{\beta\gamma}\gamma^{\beta\gamma}\gamma_\alpha
(\tilde\gamma^{12}+\tilde\gamma^{34})+\ft12g\gamma_\alpha]
\epsilon,
\eqlabel{eq:5dvar}
\end{equation}
as well as the $T^{1,1}$ variations
\begin{eqnarray}
\delta\psi_1&=&[\sqrt6\partial_{\theta_1}-\ft{i}{48}F_{\alpha\beta}
\gamma^{\alpha\beta}\tilde\gamma^1(-\tilde\gamma^{12}+\tilde\gamma^{34})
+\ft{i}2\tilde\gamma^1(1-i\tilde\gamma^{34})]\epsilon,\nonumber\\
\delta\psi_2&=&[\sqrt6\csc\theta_1\partial_{\phi_1}-\ft{i}{48}F_{\alpha\beta}
\gamma^{\alpha\beta}\tilde\gamma^2(-\tilde\gamma^{12}+\tilde\gamma^{34})
+\ft{i}2\tilde\gamma^2(1-i\tilde\gamma^{34})]\epsilon\nonumber\\
&&\qquad-\sqrt6\cot\theta_1(\partial_\psi+\ft12\tilde\gamma^{12})\epsilon,
\nonumber\\
\delta\psi_3&=&[\sqrt6\partial_{\theta_2}-\ft{i}{48}F_{\alpha\beta}
\gamma^{\alpha\beta}\tilde\gamma^3(\tilde\gamma^{12}-\tilde\gamma^{34})
+\ft{i}2\tilde\gamma^3(1-i\tilde\gamma^{12})]\epsilon,\nonumber\\
\delta\psi_4&=&[\sqrt6\csc\theta_2\partial_{\phi_2}-\ft{i}{48}F_{\alpha\beta}
\gamma^{\alpha\beta}\tilde\gamma^4(\tilde\gamma^{12}-\tilde\gamma^{34})
+\ft{i}2\tilde\gamma^4(1-i\tilde\gamma^{12})]\epsilon\nonumber\\
&&\qquad-\sqrt6\cot\theta_2(\partial_\psi+\ft12\tilde\gamma^{34})\epsilon,
\nonumber\\
\delta\psi_5&=&[3\partial_\psi-\ft1{48}F_{\alpha\beta}\gamma^{\alpha\beta}
(2-i\tilde\gamma^{12}-i\tilde\gamma^{34})
+\ft12(\tilde\gamma^{12}+\tilde\gamma^{34}+i\tilde\gamma^5)]\epsilon.
\label{eq:t11var}
\end{eqnarray}

The internal Dirac matrices provide a spinor representation of the $SO(5)$
tangent space of $T^{1,1}$.  This $\mathbf 4$ of $SO(5)$ corresponds to
taking the independent weights $i\tilde\gamma^{12}=\pm1$ and
$i\tilde\gamma^{34}=\pm1$.  Examining the variations (\ref{eq:t11var}),
we see that the Killing spinor must take the weights $i\tilde\gamma^{12}=
i\tilde\gamma^{34}=1$.  Furthermore, $\tilde\gamma^5$ is not
independent, and must have corresponding eigenvalue $\tilde\gamma^5=-1$.
Thus this $T^{1,1}$ background preserves 1/4 of the supersymmetries, and
the Killing spinor $\eta$ of $T^{1,1}$ satisfies the projections
\begin{equation}
i\tilde\gamma^{12}\eta=i\tilde\gamma^{34}\eta=-\tilde\gamma^5\eta=\eta.
\end{equation}
The $T^{1,1}$ Killing spinor equation is now trivial to solve, and the
resulting solution is
\begin{equation}
\eta=e^{\fft{i}2\psi}\eta_0,
\end{equation}
with $\eta_0$ satisfying the above projections.  This demonstrates that
the Killing spinor is charged along the $U(1)$ fiber.  This charge is of
course expected, and gives rise to a charged gravitino in five dimensions
as appropriate for the gauged supergravity theory.  Inserting this Killing
spinor into (\ref{eq:5dvar}), and rewriting this variation using curved
indices, we finally obtain
\begin{equation}
\delta\psi_\mu=[\nabla^{(5)}_\mu-\ft{i}2gA_\mu+\ft{i}{24}
(\gamma_\mu{}^{\nu\lambda}-4\delta_\mu^\nu\gamma^\lambda)F_{\nu\lambda}
+\ft12g\gamma_\mu]\epsilon,
\eqlabel{eq:n2gto}
\end{equation}
which is the appropriate gravitino variation for minimal $\caln=2$
gauged supergravity in five dimensions.

\subsection{$Y^{p,q}$ reduction of IIB}

We now turn to the case of $Y^{p,q}$.  As expected, the reduction of IIB
on $Y^{p,q}$ is almost identical to that for $T^{1,1}$.
We start with the metric on $Y^{p,q}$ written as \cite{y2,y3}
\begin{eqnarray}
ds_{10}^2&=&g_{\mu\nu}dx^\mu dx^\nu+\fft1{g^2}\Bigl[
\fft{1-y}6(d\theta^2+\sin^2\theta\,
d\phi^2)+\fft1{wv}dy^2+\fft{wv}{36}(d\beta+\cos\theta\,d\phi)^2\nonumber\\
&&+\fft19(d\psi-\cos\theta\,d\phi+y(d\beta+\cos\theta\,d\phi)+gA_{(1)})^2\Bigr],
\end{eqnarray}
where
\begin{equation}
w=\fft{2(a-y^2)}{1-y},\qquad v=\fft{a-3y^2+2y^3}{a-y^2},
\end{equation}
and introduce the natural vielbein basis
\begin{eqnarray}
&&e^1=\sqrt{\fft{1-y}6}d\theta,\qquad e^2=\sqrt{\fft{1-y}6}\sin\theta\,d\phi,
\nonumber\\
&&e^3=\fft1{\sqrt6H}dy,\qquad e^4=\fft{H}{\sqrt6}(d\beta+\cos\theta\,
d\phi),\nonumber\\
&&e^5=\fft13(d\psi-\cos\theta\,d\phi+y(d\beta+\cos\theta\,d\phi)+gA_{(1)}).
\end{eqnarray}
For later convenience, we have defined the functions
\begin{equation}
H=\sqrt{\fft{wv}6}=\sqrt{\fft{a-3y^2+2y^3}{3(1-y)}},\qquad
K=\fft{H}{2(1-y)}=\sqrt{\fft{a-3y^2+2y^3}{12(1-y)^3}},
\end{equation}
which satisfy the relation
\begin{equation}
\fft{dH}{dy}=K-\fft{y}H.
\end{equation}

In terms of the vielbeins, the $5$-form ansatz takes the same form as that
given above for the $T^{1,1}$ case.  In particular
\begin{equation}
F_{(5)}=(1+*)\left[4g\epsilon_{(5)}-\fft1{3g^2}J_{(2)}\wedge *_5F_{(2)}\right],
\eqlabel{eq:ypqf5}
\end{equation}
where the K\"ahler form is given by $J_{(2)}=e^1\wedge e^2+e^3\wedge e^4$.
Using the relations
\begin{eqnarray}
&&de^1=\sqrt6K\,e^1\wedge e^3,\qquad
de^2=\sqrt6K\,e^2\wedge e^3+\sqrt{\fft6{1-y}}\cot\theta\,e^1\wedge e^2,
\nonumber\\
&&de^3=0,\qquad
de^4=\sqrt6\left(K-\fft{y}H\right)e^3\wedge e^4-2\sqrt6K\,e^1\wedge e^2,
\nonumber\\
&&de^5=2(e^1\wedge e^2+e^3\wedge e^4)+\fft13F_{(2)},
\label{eq:ypqde}
\end{eqnarray}
we see that
\begin{equation}
d(e^1\wedge e^2)=-2\sqrt6K\,e^1\wedge e^2\wedge e^3,\qquad
d(e^3\wedge e^4)=2\sqrt6K\,e^1\wedge e^2\wedge e^3.
\end{equation}
As a result, $d(e^1\wedge e^2+e^3\wedge e^4)=0$, giving an explicit
demonstration that the K\"ahler form is closed, namely $dJ_{(2)}=0$.
As a result, the ten-dimensional Bianchi identity $dF_{(5)}=0$
again reduces to the five-dimensional Bianchi identity and equation of
motion (\ref{eq:5f}).  Similarly, the reduction of the ten-dimensional
Einstein equation follows just as in the $T^{1,1}$ case.  The resulting
bosonic fields $g_{\mu\nu}$ and $A_\mu$ are then described by the $\caln=2$
Lagrangian (\ref{eq:5lag}).

Turning to the IIB gravitino variation, we first compute the spin
connections from (\ref{eq:ypqde})
\begin{eqnarray}
&&\omega^{12}=-\sqrt{\fft6{1-y}}\cot\theta\,e^2+\sqrt6K\,e^4-e^5,\qquad
\omega^{13}=-\sqrt6K\,e^1,\qquad
\omega^{14}=\sqrt6K\,e^2,\nonumber\\
&&\omega^{23}=-\sqrt6K\,e^2,\qquad
\omega^{24}=-\sqrt6K\,e^1,\qquad
\omega^{34}=-\sqrt6\left(K-\fft{y}H\right)e^4-e^5,\nonumber\\
&&\omega^{15}=-e^2,\qquad\omega^{25}=e^1,\qquad
\omega^{35}=-e^4,\qquad\omega^{45}=e^3,
\end{eqnarray}
and
\begin{equation}
\omega^{\alpha\beta}=\omega^{(5)\,\alpha\beta}-\fft1{6g}F^{\alpha\beta}
e^5,\qquad\omega^{\alpha5}=-\fft16F^\alpha{}_\beta e^\beta.
\end{equation}
Since the 5-form reduction \eqref{eq:ypqf5} has the same form as
for the $T^{1,1}$ case, the reduction of $F\cdot\Gamma\Gamma_A$ results
in an expression identical to \eqref{eq:fdotgamma}, however with the
replacement $F_{\beta\gamma}\to-F_{\beta\gamma}$ due to the opposite
sign convention for the K\"ahler form $J_{(2)}$.

The IIB gravitino variation \eqref{eq:10var} then breaks up into the
spacetime component
\begin{equation}
\delta\psi_\alpha=[\nabla_\alpha^{(5)}-gA_\alpha\partial_\psi
+\ft{i}{12}F_{\alpha\beta}\gamma^\beta\tilde\gamma^5
+\ft1{48}F_{\beta\gamma}\gamma^{\beta\gamma}\gamma_\alpha
(\tilde\gamma^{12}+\tilde\gamma^{34})+\ft12g\gamma_\alpha]
\epsilon,
\eqlabel{eq:ypqstg}
\end{equation}
and $Y^{p,q}$ components
\begin{eqnarray}
\delta\psi_1&=&[\sqrt6(1-y)^{-1/2}\partial_\theta+\ft12\sqrt6K
\tilde\gamma^{23}(\tilde\gamma^{12}-\tilde\gamma^{34})
+\ft{i}2\tilde\gamma^1(1+i\tilde\gamma^{34})\nonumber\\
&&\qquad-\ft{i}{48}F_{\alpha\beta}
\gamma^{\alpha\beta}\tilde\gamma^1(\tilde\gamma^{12}-\tilde\gamma^{34})
]\epsilon,\nonumber\\
\delta\psi_2&=&[\sqrt6(1-y)^{-1/2}(\csc\theta\partial_\phi
-\cot\theta\partial_\beta+\cot\theta(\partial_\psi-\ft12\tilde\gamma^{12}))
-\ft12\sqrt6K\tilde\gamma^{24}(\tilde\gamma^{12}-\tilde\gamma^{34})\nonumber\\
&&\qquad+\ft{i}2\tilde\gamma^2(1+i\tilde\gamma^{34})
-\ft{i}{48}F_{\alpha\beta}\gamma^{\alpha\beta}\tilde\gamma^2
(\tilde\gamma^{12}-\tilde\gamma^{34})]\epsilon,\nonumber\\
\delta\psi_3&=&[\sqrt6H\partial_y+\ft{i}2\tilde\gamma^3(1+i\tilde\gamma^{12})
+\ft{i}{48}F_{\alpha\beta}\gamma^{\alpha\beta}\tilde\gamma^3(\tilde\gamma^{12}
-\tilde\gamma^{34})]\epsilon,\nonumber\\
\delta\psi_4&=&[\sqrt6H^{-1}\partial_\beta-\sqrt6yH^{-1}(\partial_\psi
-\ft12\tilde\gamma^{34})+\ft12\sqrt6K(\tilde\gamma^{12}-\tilde\gamma^{34})
+\ft{i}2\tilde\gamma^4(1+i\tilde\gamma^{12})\nonumber\\
&&\qquad+\ft{i}{48}F_{\alpha\beta}
\gamma^{\alpha\beta}\tilde\gamma^4(\tilde\gamma^{12}-\tilde\gamma^{34})]
\epsilon,\nonumber\\
\delta\psi_5&=&[3\partial_\psi-\ft12(\tilde\gamma^{12}+\tilde\gamma^{34}
-i\tilde\gamma^5)-\ft{i}{48}F_{\alpha\beta}\gamma^{\alpha\beta}
(\tilde\gamma^{12}+\tilde\gamma^{34}-2i)]\epsilon.
\end{eqnarray}
It is clear from the above that the Killing spinor $\eta$ on $Y^{p,q}$ is
given by
\begin{equation}
\eta=e^{\fft{i}2\psi}\eta_0,\qquad
i\tilde\gamma^{12}\eta_0=i\tilde\gamma^{34}\eta_0=\tilde\gamma^5\eta_0
=-\eta_0,
\end{equation}
and hence the spacetime component \eqref{eq:ypqstg} reduces to the expected
$\caln=2$ gravitino variation \eqref{eq:n2gto}.

Although the derivation of this reduction ansatz is straightforward (since in
essence we are only performing a Kaluza-Klein reduction along a $U(1)$
fiber), the exact Kaluza-Klein ansatz is extremely useful in that it enables
us to greatly expand the investigations of gauge theories at finite $R$-charge
chemical potential.  This is what we now turn to in the following sections.

%%%%%%%%%%%%%%%%%%%%%%%%%%%%%%%%%%%%%%%%
\section{$R$-charged black holes and their thermodynamics}

Working in the Poincar\'e patch of AdS$_5$, we may consider the following
black hole ansatz in the minimal gauged supergravity described by
\eqref{eq:5lag}:
\begin{equation}
ds_5^2=-c_1^2\,\left(dt\right)^2+c_2^2\,\left(d\vec{x}\right)^2
+c_3^2\,\left(dr\right)^2.
\eqlabel{5dmetric}
\end{equation}
Here the functions only depend on the radial coordinate $r$, namely
$c_i=c_i(r)$.  In order to work in a given $R$-charge sector, we must
also consider turning on a gauge field potential
\begin{equation}
A_t=a(r).
\eqlabel{a}
\end{equation}

Solving the equations of motion \eqref{eq:5f}, \eqref{eq:5eins} within the
ansatz \eqref{5dmetric}, \eqref{a} yields
\begin{equation}
\begin{split}
&c_1=r\ f^{1/2},\qquad c_2=r,\qquad c_3=c_1^{-1},\qquad
a'=-\frac{2Q c_1 c_3}{c_2^3},\\
&f=1-\frac{\mu}{r^4}+\frac{Q^2}{9 r^6},
\end{split}
\eqlabel{sol}
\end{equation}
where the prime denotes a derivative with respect to $r$, $\mu$ is the
non-extremality parameter and the parameter $Q$ is related to the $R$
charge of the black hole.

Let us compute the renormalized (in the sense of \cite{bk})
Euclidean gravitational action $I_E$ of \eqref{eq:5lag}.
First, we regularize \eqref{eq:5lag} by introducing
a boundary $\del \calm_5$ at fixed (large) $r$ with the
unit orthonormal space-like vector $n^\mu\propto \delta^{\mu}_r$
\begin{equation}
\begin{split}
S_5^r&=\frac {1}{16\pi G_5}\int_{\rp}^r dr \int_{\del\calm_5}d^4\xi
\sqrt{g_E}\call_E=-
\frac {1}{16\pi G_5}\int_{\rp}^r dr \int_{\del\calm_5}d^4\xi
\sqrt{-g}\call\\
&=\frac {1}{16\pi G_5}\int_{\rp}^r dr
\bigg[\frac{2c_2^2c_1c_2'}{c_3}\bigg]'\int_{\del\calm_5}d^4\xi\\
&=\frac{\b V_3}{16\pi G_5}\bigg[\frac{2c_2^2c_1c_2'}{c_3}\bigg]\bigg|_{\rp}^r\,,
\end{split}
\eqlabel{rega}
\end{equation}
where the subscript ${}_E$ indicates that all the quantities are to be
computed in Euclidean signature, and $\rp$ is the outer black hole horizon,
\ie, the largest positive root of
\begin{equation}
f(\rp)=0.
\eqlabel{rpdef}
\end{equation}
The black hole temperature and entropy are given by
\begin{equation}
\begin{split}
T_H=&\frac{\mu}{\pi \rp^3}-\frac{Q^2}{6\pi \rp^5},\qquad S_{BH}
=\frac{\rp^3 V_3}{4 G_5}.
\end{split}
\eqlabel{ts}
\end{equation}

As usual, to have a well-defined variational problem in the presence
of a boundary requires the inclusion of the Gibbons-Hawking
$S_{GH}$ term
\begin{equation}
\begin{split}
S_{GH}&=-\frac{1}{8\pi G_5}\int_{\del\calm_5}d^4\xi\sqrt{h_E}\nabla_\mu n^\mu\\
&=-\frac{\b V_3}{8\pi G_5}
\bigg[\frac{(c_1c_2^3)'}{c_3}\bigg]\,,
\end{split}
\eqlabel{sgh}
\end{equation}
where $h_{\mu\nu}$ is the induced metric on $\del\calm_5$
\begin{equation}
h_{\mu\nu}\equiv g_{\mu\nu}-n_\mu n_\nu\,.
\eqlabel{hmet}
\end{equation}
Finally, as in \cite{bk}, we supplement the combined
regularized action $\left(S_5^r+S_{GH}\right)$ by the appropriate boundary
counterterms constructed out of metric invariants on the boundary $\del\calm_5$
\begin{equation}
\begin{split}
S^{counter}&=\frac{1}{16\pi G_5}\int_{\del\calm_5}d^4\xi\,6\sqrt{h_E}\,.
\end{split}
\eqlabel{scount}
\end{equation}
In the present case the boundary curvature vanishes, so there are no extra
counterterms.  The renormalized Euclidean action $I_E$ defined as
\begin{equation}
I_E\equiv \lim_{r\to \infty}\ \biggl(
S_5^r+S_{GH}+S^{counter}\biggr),
\eqlabel{IEdef}
\end{equation}
is finite.

We now proceed to the computation of the ADM mass for the
background \eqref{5dmetric}.
Following \cite{bk} we define
\begin{equation}
M=\int_{\Sigma}d^3\xi\ \sqrt{\sigma} N_{\Sigma} \epsilon\,,
\eqlabel{massdef}
\end{equation}
where $\Sigma\equiv S^3$ is a spacelike hypersurface in $\del\calm_5$
with a timelike unit normal $u^\mu$,
$N_{\Sigma}$ is the norm of the timelike Killing vector
in \eqref{5dmetric}, ${\sigma}$ is the determinant of the induced metric
on $\Sigma$, and $\epsilon$ is the proper energy density
\begin{equation}
\epsilon=u^\mu u^\nu T_{\mu\nu}\,.
\eqlabel{epdef}
\end{equation}
The quasilocal stress tensor $T_{\mu\nu}$ for our background
is obtained from the variation of the full action
\begin{equation}
S_{tot}=S_5^r+S_{GH}+S^{counter}\,,
\eqlabel{eq:stot}
\end{equation}
with respect to the boundary metric $\delta h_{\mu\nu}$
\begin{equation}
T^{\mu\nu}=\frac{2}{\sqrt{-h}}\ \frac{\delta S_{tot}}{\delta h_{\mu\nu}}\,.
\eqlabel{qlst}
\end{equation}
An explicit computation yields
\begin{equation}
T^{\mu\nu}=\frac{1}{8\pi G_5}\biggl[
-\Theta^{\mu\nu}+\Theta h^{\mu}-3 h^{\mu\nu}\biggr]\,,
\eqlabel{tfin}
\end{equation}
where
\begin{equation}
\Theta^{\mu\nu}=\ft 12 \left(\nabla^\mu n^\nu+
\nabla^\nu n^\mu\right),\qquad \Theta=\tr \Theta^{\mu\nu}\,.
\eqlabel{thdef}
\end{equation}
Again, the renormalized stress energy tensor is finite.

Inserting the black hole solution into the above action (\ref{eq:stot}) and
renormalized stress tensor (\ref{tfin}), we obtain the action and mass
\begin{equation}
\begin{split}
I_E&=-\frac {\b\mu V_3}{16\pi G_5}\,,\\
M&=\frac{3\mu V_3}{16\pi G_5}\,.
\end{split}
\eqlabel{results}
\end{equation}
Notice that, by using \eqref{ts} and \eqref{results}, we find the
expected thermodynamical relation
\begin{equation}
I_E=\b \left(M-{\mu_{\tq}}\ \tq\right)-S_{BH} \,,
\eqlabel{thermrel}
\end{equation}
where $\mu_{\tq}$ is the chemical potential conjugate to the
physical black hole charge density
\begin{equation}
\tq=\frac{Q}{\sqrt{3}},
\eqlabel{tqdef}
\end{equation}
related to the gauge potential
$A_t$ \eqref{a} at the horizon, $r=\rp$
\begin{equation}
\mu_{\tq}=\frac{V_3}{8\pi G_5}\ \frac{A_t}{\sqrt{3}}\bigg|_{r=\rp}
=\frac{V_3}{8\pi G_5}\ \frac{Q}{\sqrt{3}\ \rp^2}\,.
\eqlabel{mutq}
\end{equation}
The factors of $\sqrt{3}$ in \eqref{tqdef} and \eqref{mutq} come from the
noncanonical normalization of the gauge field in \eqref{eq:5lag}.

We may now identify the thermodynamical quantities of the charged black hole
$\{I_E,M,S_{BH};T_{H},\mu_{\tq}\}$ with the
appropriate gauge theory quantities $\{\Omega,E,S,T,\mu_J\}$
\begin{equation}
\{I_E,M,S_{BH};T_{H},\mu_{\tq}\}
\longleftrightarrow
\{\Omega/T,E,S;T,\mu_J\}\,,
\eqlabel{translation}
\end{equation}
where the thermodynamic potential $\Omega$ is related to the
Helmholtz free energy $F$ in the standard way
\begin{equation}
\Omega=F-\mu_J J=E-T\ S -\mu_J J\,.
\eqlabel{gmF}
\end{equation}
On can explicitly verify that with the identification \eqref{translation}
the first law of thermodynamics for the grand canonical ensemble with
$\{T,\mu_J\}$ as independent variables
\begin{equation}
d\Omega= -S\ dT -J\ d \mu_J\,.
\eqlabel{1st}
\end{equation}
is satisfied automatically.  To check \eqref{1st} it is useful to use
\begin{equation}
d\left(f(\rp)\right)=0,
\eqlabel{useful}
\end{equation}
which is simply the statements that given $\{\mu,Q\}$, the radius of the
outer horizon of the BH is given by \eqref{rpdef}.

%%%%%%%%%%%%%%%%%%%%%%%%%%%%%%%%%%%%%%%%
\section{Hydrodynamics}

In this section, we examine the hydrodynamics of the $\caln=2$ Yang-Mills
plasma at finite $R$-charge which is dual to the black hole solution of
(\ref{5dmetric}), (\ref{a}) and (\ref{sol}).  In particular, using
prescription \cite{ss}, we compute the retarded Green's function of the
boundary stress-energy tensor $T_{\mu\nu}(t,x^\a)$ ($\mu=\{t,x^\a\}$) at
zero spatial momentum, and in the low-energy limit $\w\to 0$:
\begin{equation}
G^{R}_{12,12}(\w,0)=-i\int dt\,d^3x\ e^{i\w t}\theta(t)
\langle[T_{12}(t,x^\a),T_{12}(0,0)]\rangle\,.
\eqlabel{defgr}
\end{equation}
Computation of this Green's function allows a determination of the
shear viscosity $\eta$ through the Kubo relation
\begin{equation}
\eta=\lim_{\w\to 0}\frac{1}{2\w i}\left[G^A_{12,12}(\w,0)-G^R_{12,12}(\w,0)
\right]\,,
\eqlabel{kubo}
\end{equation}
where the advanced Green's function is given by
$G^A(\w,0)=\left(G^R(\w,0)\right)^\ast$.

Although the extraction of the retarded Green's function from the $R$-charged
black hole background is somewhat involved, the result turns out to be
independent of charge and chemical potential.  As demonstrated below, we find
\begin{equation}
G^{R}_{12,12}(\w,0)=-\frac{i\w s}{4\pi}\left(1+\calo\left(\frac{\w}{T}\right)
\right)\,,
\eqlabel{result}
\end{equation}
where
\begin{equation}
s=\frac{\rp^3}{4 G_5}
\eqlabel{entdef}
\end{equation}
is the Bekenstein-Hawking entropy density of the black hole.  Inserting
this expression into (\ref{kubo}) then yields the universal ratio
\begin{equation}
\frac{\eta}{s}=\frac{1}{4\pi}\,.
\eqlabel{sv}
\end{equation}

\subsection{Computation of the retarded Green's function}

We begin the computation of \eqref{defgr} by recalling that the coupling
between the boundary value of the graviton and the stress-energy tensor of
a gauge theory is given by $\delta g_2^1T_1^2/2$. According to the
gauge/gravity prescription, in order to compute the retarded thermal
two-point function \eqref{defgr}, we should add a small bulk perturbation
$\delta g_{12}(t,y)$ to the metric \eqref{5dmetric}, and compute the on-shell
action as a functional of its boundary value $\delta g_{12}^b(t)$. Symmetry
arguments \cite{ne2} guarantee that for a perturbation of this type and
metric and gauge potential of the form \eqref{5dmetric} and \eqref{a}, all
other components of a generic perturbation $\delta g_{\mu\nu}$ along with the
gauge potential perturbations $\delta A_\mu$ can be consistently set to zero.

Instead of working directly with $\delta g_{12}$, we find it convenient to
introduce the field $\p=\p(t,r)$ according to
\begin{equation}
\p=\frac 12 g^{11}\,\delta g_{12}=\frac 12 c_2^{-2}\ \delta g_{12}\,.
\eqlabel{defp}
\end{equation}
The retarded correlation function $G^R_{12,12}(\w,0)$ can be extracted
from the (quadratic) boundary effective action $S_{boundary}$ for the metric
fluctuations $\p^b$ given by
\begin{equation}
S_{boundary}[\p^b]=\int \frac{d^4k}{(2\pi)^4}\,\p^{b}(-\w)\,\calf(\w,r)\,
\p^{b}(\w)\bigg|^{\del\calm_5}_{horizon}\,,
\eqlabel{sssb}
\end{equation}
where
\begin{equation}
\p^b(\w)=\int \frac{d^4k}{(2\pi)^4} e^{-i\w t}\
\p(t,r)\bigg|_{\del\calm_5}\,.
\eqlabel{pb}
\end{equation}
In particular, the Green's function is given simply by
\begin{equation}
G^R_{12,12}(\w,0)=\lim_{\del\calm_5^r\to\del\calm_5}\ 2\ \calf^r(\w,r)\,,
\eqlabel{Gr}
\end{equation}
where $\mathcal F$ is the kernel of (\ref{sssb}).
The boundary metric functional is defined as
\begin{equation}
S_{boundary}[\p^b]=\lim_{\del\calm_5^r\to\del\calm_5}\biggl(
S^r_{bulk}[\p]+S_{GH}[\p]+S^{counter}[\p]\biggr)\,,
\eqlabel{bounfu}
\end{equation}
where $S^r_{bulk}$ is the bulk Minkowski-space effective supergravity action
\eqref{eq:5lag} on a cut-off space $\calm_5^r$ (where $\calm_5$ in
\eqref{5dmetric} is regularized by the compact manifold $\calm_5^r$ with a
boundary $\del\calm_5^r$).  Also, $S_{GH}$ is the standard Gibbons-Hawking
term over the regularized boundary $\del\calm_5^r$.  The regularized bulk
action $S^r_{bulk}$ is evaluated on-shell for the bulk metric fluctuations
$\p(t,r)$ subject to the following boundary conditions:
\begin{equation}
\begin{split}
&(a):\quad\lim_{\del\calm_5^r\to\del\calm_5} \p(t,r)=\p^b(t)\,,\\
&(b):\quad\p(t,r)\ \hbox{is an incoming wave at the horizon}\,.
\end{split}
\eqlabel{bc}
\end{equation}
The purpose of the boundary counterterm $S^{counter}$ is to remove
divergent (as $\del\calm_5^r\to\del\calm_5$) and $\w$-independent
contributions from the kernel $\calf$ of \eqref{sssb}.

We find that the effective bulk action for $\p(t,r)$ in the
supergravity background \eqref{5dmetric}, \eqref{a} takes the form
\begin{equation}
\begin{split}
S_{bulk}[\p]\equiv&\frac{1}{16\pi G_5}\int d^{5}x\ \call_5
=\frac{1}{16\pi G_5}\int d^{5}x\ \biggl[\\
&c_1c_2^3c_3
\biggl\{\frac{1}{2c_1^2} \left(\del_t \p\right)^2-\frac{1}{2c_3^2}
\left(\del_r\p\right)^2\biggr\}\\
&+\biggl\{-\del_t\left(\frac{2c_2^3c_3}{c_1}\ \phi\del_t\phi\right)
+\del_r\left(
\frac{2c_2^3c_1}{c_3}\ \phi\del_r\phi+\frac{c_1c_2^2c_2'}{c_3}\,\phi^2
\right)
\biggr\}
\biggr]\,.
\end{split}
\eqlabel{ac2}
\end{equation}
The second line in \eqref{ac2} is the effective action for a minimally coupled
scalar in the geometry \eqref{5dmetric}, while the third line is a total
derivative. Thus the bulk equation of motion for $\p$ is that of a minimally
coupled scalar in \eqref{5dmetric}.  The latter equation is simplified by
introducing a new radial coordinate%
\footnote{With such a definition, $x$ has a range $x\in[0,1]$ with
$x\to 0_+$ being the horizon and $x\to 1_-$ the boundary.}
\begin{equation}
x\equiv \frac{c_1}{c_2},
\eqlabel{xdef}
\end{equation}
which using \eqref{sol} can be inverted to give the near horizon expansion
\begin{equation}
c_2(x)=\rp-\frac{9\rp^7}{2(Q^2-18\rp^6)}\ x^2+\frac{81\rp^{13}(11Q^2-90\rp^6)}
{8(Q^2-18\rp^6)^3}\ x^4+\calo(x^6).
\eqlabel{c2x}
\end{equation}
Decomposing $\p$ as
\begin{equation}
\p(t,x)=e^{-i\w t}\p_\w(x)\,,
\end{equation}
we find that the equation of motion reduces to
\begin{equation}
\begin{split}
0=&\p_\w''-\frac{(18 c_2^8-c_2^2 Q^2+6 x c_2'c_2 Q^2+12 x^2
\left(c_2'\right)^2 Q^2)}{x c_2^2 (Q^2-18 c_2^6)}\,\p_\w'
-\frac{9 \w^2 c_2^2 c_2' (c_2+2 c_2' x)}{x^3 (Q^2-18 c_2^6)} \,\p_w,
\end{split}
\eqlabel{eomp}
\end{equation}
where primes denote derivatives with respect to $x$.

A low-frequency solution of \eqref{eomp} which is an incoming wave at the
horizon, and which near the boundary satisfies
\begin{equation}
\lim_{x\to 1_-}\ \p_\w(x)=1\,,
\eqlabel{blim}
\end{equation}
can be written as
\begin{equation}
\p_\w(x)=x^{-i\ww }\ \biggl(F_0(x)+i\ww\ F_{\w}(x)+\calo(\ww^2)\biggr)\,,
\eqlabel{taylorsol}
\end{equation}
where $\ww$
\begin{equation}
\ww=\frac{\w}{2\pi T}\,.
\eqlabel{defq}
\end{equation}
The functions $\{F_0,\ F_{\w}\}$, which are smooth at the horizon, satisfy
the following differential equations:
\begin{equation}
\begin{split}
0=&F_0''-\frac{(18 c_2^8-c_2^2 Q^2+6 x c_2'c_2 Q^2+12 x^2 \left(c_2'\right)^2
Q^2)}{x c_2^2 (Q^2-18 c_2^6)}\,F_0',\\
0=&F_\w''-\frac{(18 c_2^8-c_2^2 Q^2+6 x c_2' c_2 Q^2 +12 x^2
\left(c_2'\right)^2 Q^2)}{x c_2^2 (Q^2-18 c_2^6)}\,F_\w'
-\frac 2x\,F_0'+\frac{6c_2'Q^2(c_2+2 x c_2')}{x c_2^2(Q^2-18 c_2^6)}\,F_0\,.
\end{split}
\eqlabel{difff}
\end{equation}
Notice that for $Q=0$ the dependence on the specific background geometry
($c_2(x)$ dependence) drops out and \eqref{difff} coincides with universal
equations derived in \cite{u3}.  In the present case, however, since the $Q$
dependence appears highly non-trivial, universality is far from assured.

The general solution of the first equation in \eqref{difff} takes the form
\begin{equation}
\begin{split}
F_0=\calc_1+\calc_2\int dx\,{\exp}\biggl\{
\int^xdy\,\frac{18 c_2^8-c_2^2 Q^2+6 y c_2'c_2 Q^2+12 y^2 \left(c_2'\right)^2
Q^2}{y c_2^2 (Q^2-18 c_2^6)}
\biggr\},
\end{split}
\eqlabel{difff1}
\end{equation}
where $\calc_i$ are integration constants.
Using the near horizon expansion \eqref{c2x}, we find as $x\to 0_+$
\begin{equation}
F_0=\calc_1+\calc_2\left(\ln x-\frac{27Q^2\rp^6}{2(Q^2-18\rp^6)^2}\ x^2
+\calo\left(x^4\right)\right).
\eqlabel{smallx}
\end{equation}
Thus nonsingularity of $F_0$ at the horizon requires $\calc_2=0$. The
boundary condition \eqref{blim} further specifies
\begin{equation}
F_0(x)=1.
\eqlabel{f0f}
\end{equation}
The solution of the second equation in \eqref{difff} is a bit more complicated
and will be discussed in the Appendix.  We note here that unlike the universal
case \cite{u3}, given \eqref{f0f} and the boundary conditions, this solution
is non-trivial. Altogether we have
\begin{equation}
\p(t,x)=e^{-i\w t}\ x^{-i\ww}\ \left(1+i\ww F_\w(x)+\calo(\ww^2)\right)\,.
\eqlabel{solff}
\end{equation}

Once the bulk fluctuations are on-shell (\ie,\ satisfy equations of motion)
the bulk gravitational Lagrangian becomes a total derivative. From \eqref{ac2}
we find (without dropping any terms)
\begin{equation}
\call_{5}=\del_t J^t+\del_r J^r,
\eqlabel{tder}
\end{equation}
where
\begin{equation}
\begin{split}
J^t=&-\frac{3c_2^3c_3}{2c_1}\,\phi\del_t\phi,\\
J^r=&\frac{3c_2^3c_1}{2c_3}\,\phi\del_r\phi+\frac{c_2^2c_1c_2'}{c_3}\,\phi^2.
\end{split}
\eqlabel{jr}
\end{equation}
Additionally, the Gibbons-Hawking term provides an extra contribution so that
\begin{equation}
J^r\to J^r-\frac{2c_2^3c_1}{c_3}\,\phi\del_r\phi-\frac{(c_1c_2^3)'}{c_3}\,\p^2.
\eqlabel{ghshift}
\end{equation}

We are now ready to extract the kernel $\calf$ of \eqref{sssb}.
The regularized boundary effective action for $\p$ is
\begin{equation}
\begin{split}
S_{boundary}[\p]^r
=&\frac{1}{16\pi G_5}\int_{\del\calm^r_5}dt\,d^3x\,
\biggl(-\frac{c_2^3c_1}{2c_3}\,\phi\del_r\phi\biggr)+c.t.,
\end{split}
\eqlabel{breg}
\end{equation}
where as prescribed in \cite{ss}, we need only keep the boundary contribution.
In \eqref{breg} $c.t.$ stands for (finite) contact terms that will not be
important for computations.  Substituting \eqref{solff} into \eqref{breg} we
can obtain $\calf^r(\w,r)$:
\begin{equation}
\begin{split}
\calf^r(\w,r)=&-\frac{i\ww}{32\pi G_5}\left(1+\calo\left(\ww\right)\right)
\,\times\,\frac{c_2^3(r)c_1(r)}{c_3(r)}\left(\frac{c_1(r)}{c_2(r)}\right)'
\times \biggl\{1-\frac{dF_\w(x)}{dx}\biggr\},
\end{split}
\eqlabel{regk}
\end{equation}
where we have recalled the definition of $x$ in \eqref{xdef}.
Thus, to order $\calo(\ww^2)$, we have
\begin{equation}
\begin{split}
\lim_{r\to \infty}
\calf^r(\w,r)=&-\frac{i\ww\mu }{32\pi G_5}\
\times\ \lim_{x\to 1_-} \left\{1-\frac{dF_\w(x)}{dx}\right\},
\end{split}
\eqlabel{kenfin}
\end{equation}
where we used \eqref{sol}.
Now, given \eqref{fder} and \eqref{finf}, we finally obtain
\begin{equation}
\begin{split}
\lim_{r\to \infty}
\calf^r(\w,r)=&\frac{i\ww\mu }{32\pi G_5}\
\times\ \lim_{x\to 1_-} F(x)\\
=&\frac{i\ww\mu }{16\pi G_5} \left(-\frac{\pi T\rp^3}{\mu}\right)
=-\frac{i\w}{8\pi}\ \frac{\rp^3}{4G_5}
=-\frac{i\w s}{8\pi},
\end{split}
\eqlabel{Gr1}
\end{equation}
where we have used the expression for the entropy density \eqref{entdef} as
well as the definition \eqref{defq}.  Using \eqref{Gr} to extract the
Green's function from $\calf^r$ then gives
\begin{equation}
G_{12,12}^R(\omega,0)\approx-\fft{i\w s}{4\pi},
\end{equation}
at least in the low frequency limit $\omega\to0$.  This is the result
claimed in \eqref{result}, giving rise to the universal ratio of
shear viscosity to entropy density \eqref{sv}.

\subsection{The sound speed}

We conclude this section by computing the speed of sound in the SCFT$^{p,q}$
plasma.  Using the fact that the thermodynamic potential is
\begin{equation}
\Omega=-P V_3
\eqlabel{pressure}
\end{equation}
where $P$ is the pressure, we find from \eqref{translation} and \eqref{results}
that
\begin{equation}
P=\frac 13 \,\frac{E}{V_3},
\eqlabel{pe}
\end{equation}
which further implies that the speed of sound is
\begin{equation}
c_s^2=\frac {1}{V_3}\ \frac{\del P}{\del E}=\frac 13,
\end{equation}
independent of the value for the chemical potential.

%%%%%%%%%%%%%%%%%%%%%%%%%%%%%%%%%%%%%%%%
\section{Conclusion}

In the first half of this paper, we have derived explicit Kaluza-Klein
reduction ans\"atze of IIB on $T^{1,1}\equiv Y^{1,0}$ and $Y^{p,q}$,
yielding in both cases minimal $\caln=2$ gauged supergravity in five dimensions.
Although the Kaluza-Klein spectra of these reductions include additional
vector multiplets in the massless sector ($SU(2)\times SU(2)$ for $T^{1,1}$
or $SU(2)\times U(1)$ for generic $Y^{p,q}$), these vectors cannot be
retained in a consistent truncation \cite{hmp}.  In fact, this inconsistency
even precludes the retention of the vectors in the $U(1)^2$ subgroups of
the above groups.  As a result, we see that it is in fact not possible to
realize the $\caln=2$ STU model from $T^{1,1}$ or $Y^{p,q}$ reduction, even
though this attractive possibility is otherwise suggested from the linearized
Kaluza-Klein analysis.

Our main result is a demonstration that the shear viscosity of SCFT$^{p,q}$
plasma with nonzero $U(1)_R$ symmetry charge chemical potential is universally
related to the entropy density \eqref{univ}.  For $p=q$, a cone over $Y^{p,p}$
is a $\zet_{2p}$ orbifold of $\complex^3$, and hence the dual superconformal
quiver plasma is just a $\zet_{2p}$ orbifold of the $\caln=4$ Yang-Mills theory.
It is probably not surprising that the orbifold quivers of $\caln=4$ Yang-Mills
plasma have a universal shear viscosity with nonzero chemical potential,
much like the parent gauge theory \cite{r1,r2,r3,r4}.  What is rather
unexpected, however, is that the universality \eqref{univ} is also true for
the $Y^{1,0}\equiv T^{1,1}$ superconformal gauge quiver, which arises as a
nontrivial
superconformal infrared fixed point of the renormalization group flow from the
$\zet_2$ orbifold of $\caln=4$ supersymmetric Yang-Mills in the ultraviolet
\cite{kw}.  Thus it is natural to conjecture that the ratio ${\eta}/{s}$ is a
constant along the renormalization group flow, and as such, must be true for
any strongly coupled gauge theory plasma with nonzero chemical potential which
allows for a (gauged) supergravity dual. Needless to say, it would be very
interesting to prove this conjecture, and if it is true, understand its
applications for the charged black holes in string theory. In particular,
one should try to understand the shear viscosity of the gauge theory plasma
with a finite chemical potential but with $c_s\ne 1/\sqrt{3}$.
An example of exactly such a model would be the cascading gauge theory
plasma \cite{bh1,bh2,bh3,bh4}.

\section*{Acknowledgments}
We would like to thank Rob Myers, Andrei Starinets and Leo Pando Zayas for
valuable discussions.  AB would like to thank the Benasque Center for Science
and the European Centre for Theoretical Studies in Nuclear Physics
and Related Areas for hospitality during the final stages of this work.
JTL would like to thank the National Center for Theoretical Sciences
(Taiwan) and the National Taiwan University Department of Physics for
hospitality.  AB research at Perimeter Institute is supported in part by the
Government of Canada through NSERC and by the Province of Ontario through MEDT.
AB gratefully acknowledges further support by an NSERC Discovery grant.  JTL is
supported in party by the US Department of Energy under grant DE-FG02-95ER40899.

%%%%%%%%%%%%%%%%%%%%%%%%%%%%%%%%%%%%%%%%
\appendix
\section{Solution for $F_\w$}
As is clear from \eqref{kenfin}, what we really need for the computation of
the shear viscosity is not $F_\w$, but rather it's derivative.  We thus
introduce
\begin{equation}
F(x)=-\frac 1x+\frac{dF_\w(x)}{dx}.
\eqlabel{fder}
\end{equation}
Given the analytical solution for $F(x)$, $F_\w$ can be found by integrating
\eqref{fder}
\begin{equation}
F_\w=-\int_x^1 dy\, F(y)+\ln x.
\eqlabel{fder1}
\end{equation}
We now note that nonsingularity of $F_w$ at the horizon $x\to 0_+$ implies that
\begin{equation}
F=\frac 1x+\calo(1).
\eqlabel{fbc}
\end{equation}
Using the definition \eqref{fder}, one finds from the second equation in
\eqref{difff}
\begin{equation}
0=\frac{dF}{dx}-\frac{18 c_2^8-c_2^2 Q^2+6xc_2 c_2'Q^2+12x^2
\left(c_2'\right)^2Q^2}{x c_2^2(Q^2-18c_2^6)}\,F.
\eqlabel{feqx}
\end{equation}
In terms of the $r$ coordinate [see \eqref{xdef}] \eqref{feqx} takes form
\begin{equation}
0=\frac{dF}{dr}-\frac{(5 Q^4+18 r^6 Q^2-54 Q^2 \mu r^2+108 \mu^2 r^4)}
{r (Q^2-6 \mu r^2) (9 r^6+Q^2-9 \mu r^2)}\,F,
\eqlabel{feqr}
\end{equation}
where we used the explicit background solution \eqref{sol}.
A general solution of \eqref{feqr} takes the form
\begin{equation}
F=\frac{\calc r^5}{(9r^6+Q^2-9\mu r^2)^{1/2}(Q^2-6\mu r^2)},
\eqlabel{fsolr}
\end{equation}
where $\calc$ is an arbitrary integration constant.
Using $r=r(x)=c_2(x)$ given by \eqref{c2x}, the boundary condition
\eqref{fder1} determines
\begin{equation}
\calc=\frac{18\rp^6-Q^2}{\rp^2}=18\pi T\rp^3,
\eqlabel{ccomp}
\end{equation}
where we used the expression for the temperature \eqref{ts} and \eqref{rpdef}.
Using \eqref{fsolr} and \eqref{ccomp}, we can finally evaluate
\begin{equation}
\lim_{x\to 1_-}F(x)=\lim_{r\to \infty}F(r)=-\frac{\calc}{18\mu}
=-\frac{\pi T\rp^3}{\mu}.
\eqlabel{finf}
\end{equation}
%

%%%%%%%%%%%%%%%%%%%%%%%%%%%%%%%%%%%%%%%%

\end{document}